\documentclass[12pt,preprint]{aastex}

\def\gs{\mathrel{\raise0.35ex\hbox{$\scriptstyle >$}\kern-0.6em
\lower0.40ex\hbox{{$\scriptstyle \sim$}}}}
\def\ls{\mathrel{\raise0.35ex\hbox{$\scriptstyle <$}\kern-0.6em
\lower0.40ex\hbox{{$\scriptstyle \sim$}}}}

\shorttitle{CI J=1--0 emission as H$_2$ gas mas tracer}
\shortauthors{Papadopoulos, \& Greve}

\begin{document}

\title{CI  emission  in  Ultra   Luminous  Infrared  Galaxies as a
molecular gas mass tracer}

\author{Padeli \ P.\ Papadopoulos}
\affil{Institut f\"ur Astronomie, ETH Zurich, 8093 Z\"urich, Switzerland}
\email{papadop@phys.ethz.ch}

\and 

\author{Thomas\  R. Greve\altaffilmark{1}
  \affil{California Institute of Technology, Pasadena, CA 91125, USA }}
\email{tgreve@submm.caltech.edu}

\altaffiltext{1}{also Institute for  Astronomy,  University of
Edinburgh, Blackford Hill, Edinburgh EH9 3HJ, UK}

\begin{abstract}

We present new sensitive  wide-band measurements of the fine structure
line $  ^3 P _1\rightarrow $  $^3$$P _0$ (J=1--0, 492  GHz) of neutral
atomic carbon (CI) in the two typical Ultra Luminous Infrared Galaxies
NGC 6240  and Arp 220.  We then  use them along with  several other CI
measurements in  similar objects found  in the literature  to estimate
their  global molecular  gas content  under the  assumption of  a full
CI-H$_2$ concomitance.  We find  excellent agreement between the H$_2$
gas mass estimated  with this method and the  standard methods using $
^{12}$CO.  This  may provide a  new way to  measure H$_2$ gas  mass in
galaxies, and one  which may be very valuable in  ULIRGs since in such
systems  the bright  $ ^{12}$CO  emission is  known  to systematically
overestimate  the  gas  mass  while  their  $  ^{13}$CO  emission  (an
often-used alternative) is usually very weak.  At redshifts $\rm z\geq
1$ the CI J=1-0 line shifts to much more favorable atmospheric windows
and can  become a viable alternative  tracer of the  H$_2$ gas fueling
starburst events in the distant Universe.

\end{abstract}


\section{Introduction}

The CI  J=1--0 line emission  emerging from molecular clouds  has been
detected early  (Phillips \& Huggins  1981), but its  interpetation as
emanating from only a narrow  CII/CI/CO transition zone at the surface
of  these  clouds  (e.g.   Tielens  \& Hollenbach  1985a,b)  seems  to
preclude its use as a bulk  H$_2$ gas tracer, except perhaps to derive
lower  limits.  Nevertheless,  despite  the considerable  difficulties
involved  in  imaging  wide  areas   in  CI  emission  (caused  by  an
unfavorable  atmospheric window  at 492  GHz, and  the small  beams of
single-pixel detectors  mounted on large  aperture sub-mm telescopes),
early work has  shown CI to be widely  distributed and well-correlated
with  other H$_2$ gas  mass tracers  such as  $ ^{13}$CO  (Keene 1985;
Keene  1997 and  references therein).   Further  observational efforts
have confirmed  this CO-CI  concomitance in molecular  clouds together
with a remarkably constant $\rm  N(C)/N(CI)$ ratio over the {\it bulk}
of the  CO-bright gas (Ojha et al.   2001; Ikeda et al.   2002). A new
comprehensive  study  has shown  how  this  could  be established  and
advocated the use  of CI line emission as a  molecular gas mass tracer
in galaxies (Papadopoulos, Thi, \& Viti 2004).

Ultra  Luminous  Infrared  Galaxies  (ULIRGs) are  powered  mainly  by
 starburst events  (Genzel et al.   1998) that yield  $\rm L_{FIR}\geq
 10^{12}  L_{\odot }$  making them  the most  luminous objects  in the
 local Universe  (Sanders et al.   1988; Sanders \& Mirabel  1996).  A
 number of  reasons make them a good  sample to test CI  emission as a
 bulk H$_2$ gas mass tracer in galaxies, namely:

\noindent
a)  They  are gas-rich  systems  but  with  their H$_2$  distributions
 usually so compact (e.g.  Downes  \& Solomon 1998; Bryant \& Scoville
 1999) that single pointings  with single-dish telescopes are adequate
 to record all  their CI/CO flux density (of  prime importance for the
 difficult CI observations).

\noindent
b) Their H$_2$ gas mass has been estimated using several methods other
than  the standard  one based  solely on  the $  ^{12}$CO  J=1--0 line
luminosity (e.g.  Solomon et al.   1997; Downes \& Solomon 1998; Glenn
\& Hunter  2001; Lihong et al.  2003).  Thus they  constitute an ideal
test-bed for any new H$_2$ mass-tracing method.

\noindent
c)  In  these  merger/starburst  systems  intense  FUV  radiation  and
powerful  tidal  fields  drive   a  two-phase  differentation  of  the
molecular gas  (Aalto et al.  1995).  The  diffuse, $ ^{12}$CO-bright,
and non-virialized  intercloud phase  is probably responsible  for the
systematic overestimate of H$_2$ gas  mass by factors of $\sim 4-5$ in
starburst environments (Solomon et al.  1997; Downes \& Solomon 1998).
It is thus interesting to see  how a new method of measuring H$_2$ gas
mass will fare under such conditions.

\noindent
d) Early  applications of  this method in  Arp 220 (Gerin  \& Phillips
1998),  and the  Cloverleaf QSO  at z=  2.5 (Weiss  et al.  2003) have
yielded  an H$_2$ mass  that is  in good  agreement with  the standard
methods that use $ ^{12}$CO and dust continuum emission.

\noindent
e) ULIRGs, as possible precursors of optically-bright QSOs (Sanders et
al.   1988; Tacconi  et al.   2002), or  as the  nearby  templates for
starburst, dust-enshrouded galaxies observed at high redshifts (Smail,
Ivison \& Blain 1997; Hughes et al.  1998), are interesting systems in
their own right.  Molecular gas being the ``fuel'' of their prodigious
star-forming  activity makes  any new  method of  estimating  its mass
particularly valuable.

In this  work we present new  sensitive observations of  the CI J=1--0
line in two prominent ULIRGs namely  Arp 220 and NGC 6240, and collect
available  CI measurements  in  similar objects  from the  literature,
which are  then  used  to   estimate  their  H$_2$  gas  mass  content.
Throughout  this paper  we adopt  an $\rm  H_{\circ}=75\ km  \ s^{-1}\
Mpc^{-1} $ and $\rm q_{\circ}=1/2$.

\section{Observations and results}

We used the 15-meter James Clerk Maxwel Telescope (JCMT) on Mauna Kea,
Hawaii,  to  observe the  $  ^3P_1\rightarrow  ^3$$P_0$ (J=1--0)  fine
structure line at $\rm \nu  _{rest}=492.160$ GHz in the ULIRGs Arp~220
(January  3nd 2003)  and  NGC  6240 (February  27th  1999).  Cold  and
extremely  dry  weather  with   $\tau  _{225  GHz}\la  0.035$  ensured
excellent  conditions  for  high-frequency observations  with  overall
system temperatures  of the  SSB-tuned W-receiver of  $\rm T_{sys}\sim
(700-900)$  K.  Rapid beam-switching  at a  chop-throw of  30$''$ (Az)
with a frequency of 1 Hz produced stable baselines, and the use of the
Digital Autocorrelation  Spectrometer (DAS) at its widest  mode of 1.8
GHz ($\sim 1100\rm\  km\ s^{-1} $) enabled the  full velocity coverage
of the wide extragalactic lines often observed in such galaxies.  This
is particularly important since e.g.  Arp 220 has a $\rm FWHM\sim 500\
km\  s^{-1}$  and in  the  past  multiple  frequency settings  of  the
spectrometer were  necessary in  order to cover  the entire  CI J=1--0
line (Gerin  \& Phillips 1998).  The aperture  efficiency was measured
repeatedly  using  Mars,  an   important  task  in  sub-mm  telescopes
operating at such high  frequencies since residual thermal distortions
and  mechanical  deformations  of  the  dish  can  then  affect  their
performance significantly.   We deduced  $\langle \eta _{a}  \rangle =
0.35$,  lower than  the one  quoted  in JCMT  Observer's Guide  ($\eta
_a=0.45  $),  but within  the  expected  variations (Friberg,  private
communication).  The beam size  at these frequencies (used to estimate
the  aforementioned   efficiencies)  is  $\rm   \theta  _{HPBW}=10''$.
Frequent pointing checks showed the pointing error to be $\sim 3.5''$.

 Frequent  observations  of  strong  spectral-line sources  yielded  a
calibration uncertainty  of $\sim 20\%  $, which along with  the $\sim
25\%$ uncertainty of adopted value of aperture efficiency dominate the
error  in  the reported  velocity-integrated  line  fluxes. These  are
estimated from

\begin{equation}
\rm S_{CI}=\int  _{\Delta V}  S_{\nu } dv  = \frac{8 k_B}{\eta  _a \pi
D^2} K_c (x)\int  _{\Delta V} T^* _A dv=  \frac{15.6 (Jy/K)}{\eta _a} K_c(x)
\int _{\Delta V} T^* _A dv,
\end{equation}

\noindent
where $\rm K_c  (x) =x^2/(1-e^{-x^2}), x=\theta _s/(1.2\theta _{HPBW})
$ ($\theta _s$=source diameter) accounts for the geometric coupling of
the gaussian part  of the beam with a  finite-sized, disk-like source.
For NGC 6240 it is $\rm K_c(x)=1.15$, while for all other sources $\rm
K_c(x)\sim 1$.

The  CI spectra  are shown  in  Figures 1,  2 along  with overlayed  $
^{12}$CO and $ ^{13}$CO  J=2--1 spectral lines.  The correspondance is
excellent, while the weakness of  the observed $ ^{13}$CO emission (an
often-used alternative  to the much  more optically thick  $ ^{12}$CO)
elevates CI  emission (if  indeed fully concomitant  with H$_2$)  to a
good alternative H$_2$ gas tracer in ULIRGs.

\section{Global molecular gas mass estimates using the CI line emission}

We use  the measured $\rm S_{CI}$  to deduce the  global molecular gas
mass from

\begin{equation}
\rm \frac{M_{CI}(H_2)}{M_{\odot}}=4.92\times 10^{10} h ^{' -2}
\frac{(1+z-\sqrt{1+z})^2}{1+z}
\left[\frac{X_{CI}}{10^{-5}}\right]^{-1}
\rm \left[\frac{A_{10}}{10^{-7} s^{-1}}\right]^{-1} Q_{10} ^{-1}
\left[ \frac{S_{CI}}{Jy\ km\ s^{-1}}\right],
\end{equation}

\noindent
 where   $\rm   h^{'}=0.75    $,   the   Einstein   coefficient   $\rm
A_{10}=7.93\times  10^{-8} s^{-1} $,  and $\rm  Q_{10}=Q_{10}(n, T_k)$
depends  on the  gas excitation  conditions (see  Papadopoulos  et al.
2004).    The  $\rm   [CI]/[H_2]$  abundance   chosen  here   is  $\rm
X_{CI}=3\times  10^{-5} $,  identical to  that  used by  Weiss et  al.
(2003) in their H$_2$ mass estimate  in the Cloverleaf.  This value is
the average between the minimum found  for the bulk of the Orion A and
B clouds  ($\sim 10^{-5}$;  Ikeda et al.   2002) and in  the starburst
environment of  the nucleus of  M82 ($\sim 5\times 10^{-5}$;  White et
al.  1994) (assuming $\rm [CO]/[H_2]=10^{-4}$).  The results are shown
in Table 1  where $\rm M_{CI}(H_2)$ can be  directly compared with the
best  available $\rm  M_{CO}(H_2)$ estimates.   In the  cases  where a
range of physical conditions is reported for the gas, the average $\rm
Q_{10}$ excitation factor is adopted.

  The agreement  is remarkable,  especially given that  the CO-derived
H$_2$ masses reported  here are based on much  better methods than the
mere  application of  the Galactic  $\rm  X$ factor  (which in  ULIRGs
overestimates H$_2$  mass by a factor  of $\sim 5$). It  must be noted
though that even such methods leave uncertainties of a factor of $\sim
2$ in their  $\rm M(H_2)$ estimates (e.g.  see  Tacconi et al.  1999).
In    the    one   case    of    IRAS    10565+2448    it   is    $\rm
M_{CI}(H_2)/M_{CO}(H_2)\sim 4$, possibly too  large a difference to be
attributed solely to the  aforementioned uncertainties of the CO-based
methods.   In starburst  environments higher  $\rm X_{CI}$  values are
possibly more  appropriate (e.g.  Schilke et al.   1993 and references
therein). Indeed,  assuming CI as  abundant as in  the center of  M 82
would keep  the CI  and CO-deduced H$_2$  mass estimates  in agreement
within factors of  $\sim 2$, while bringing $\rm  M_{CI}(H_2)$ in IRAS
10565+2448 in better accord to the CO-deduced value.  Nevertheless the
latter  case highlights  an {\it  irreducible} uncertainty  when using
optically thin emission from {\it  any} molecular or atomic species to
trace H$_2$ gas mass.

  It is  important to  note that unlike  quiescent spirals  ULIRGs can
have $\ga  50\%$ of  their H$_2$ mass  at densities  $\rm n(H_2)>10^4\
cm^{-3}$ (Gao  \& Solomon 2004), for which  one-dimensional static PDR
models predict  much lower $\rm X_{CI}$  values, effectively rendering
CI emission  incapable of tracing its  mass. This does not  seem to be
happening,  lending further  support to  a full  CI-H$_2$ concomitance
over  most   of  the  parameter  space   characterizing  the  physical
conditions of  H$_2$ in  galaxies.  Furthermore CI  emission may  be a
good alternative  molecular gas tracer particularly  in ULIRGs because
the often-used optically thin $  ^{13}$CO emission is very weak, while
$ ^{12}$CO J=1--0 and  a Galactic conversion factor overestimates $\rm
M(H_2)$ because  of the  presence of non-virialized  $ ^{12}$CO-bright
H$_2$ gas.   Apart from  the uncertainty of  the assumed  $\rm X_{CI}$
(which plagues also the $ ^{13}$CO measurements since C and $ ^{13}$CO
emerge from tightly coupled chemical  routes), the CI J=1--0 line is a
more  straightforward H$_2$  mass  tracer  than CO  due  to a  simpler
partition  function and  a  significantly reduced  sensitivity to  the
ambient gas excitation conditions.

It can be argued that  a proper comparison between the various tracers
of molecular gas must be between transitions within the same frequency
domain  because  of  the  significant variations  in  the  atmospheric
absorption (and  thus in  the resulting effective  system sensitivity)
across the mm/sub-mm wavelengths.  For  CI and CO this is presented by
Papadopoulos et al. (2004) where it is shown that while the $ ^{12}$CO
J=4--3  (461 GHz)  transition  can  be indeed  much  stronger than  CI
J=1--0,  the  latter  offers  decisive advantages  for  cooler  and/or
diffuse gas ($\rm T_k\la 20\ K , n< 10^4\ cm^{-3}$) where the emission
of the former is significantly  surpressed.  In ULIRGs the very weak $
^{13}$CO  emission  makes  the   much  brighter  CI  J=1--0  a  better
alternative H$_2$ mass tracer when using the same telescope to observe
unresolved sources.  This of  course assumes CI observations conducted
during  suitable  dry  weather   conditions,  which  occur  much  less
frequently   than  the   those   needed  for   sensitive  $   ^{13}$CO
observations. However this situation is expected to change drastically
when the new generation of mm/sub-mm single dish telescopes (APEX) and
interferometer arrays (ALMA) become operational in the excellent sites
in  the   Atacama  Desert  plateau  in  Northern   Chile.   There  dry
atmospheric  conditions,  suitable  for  CI J=1--0  observations,  are
expected for  $\sim 50\%$  of the time  and with much  less pronounced
diurnal variations (Radford \& Nyman 2001 and references therein).

More CI J=1--0  observations of ULIRGs are urgently  needed to further
test the notion of CI as a bulk H$_2$ gas mass tracer.  Application of
this  technique  to  quiescent   nearby  galaxies  is  also  of  prime
importance but then extensive CI imaging may be needed to match the CO
and  CI imaged  areas. Finally  CI  observations of  galaxies at  high
redshifts with known CO  detections can further test the effectiveness
of this method while circumventing  the problem of the low atmospheric
transmission at 492~GHz (J=1--0) and 809~GHz (J=2--1).

\section{Conclusions}

We report on new sensitive CI J=1--0 measurements in the two prominent
ULIRGs NGC~6240 and Arp 220, and collect CI measurements available for
similar objects  from the literature. We  then use them  to derive the
molecular gas mass  under the assumption of fully  concomitant CI, CO,
and H$_2$ and find very good agreement between the CI and the CO-based
estimates within the uncertainties  of the two methods.  The faintness
of the $  ^{13}$CO emission (often used instead  of the more optically
thick $ ^{12}$CO)  in these two galaxies and  in similar systems makes
CI emission  a good  alternative tracer of  their molecular  gas mass.
However  given  the  much  stricter  constraints  on  the  atmospheric
conditions  needed for  sensitive CI  observations (very  dry weather)
than those  for $ ^{13}$CO, the  practical advantages of CI  as a good
optically thin H$_2$  gas mass tracer will be  fully realised when the
new  generation  of mm/sub-mm  single-dish  (APEX) and  interferometer
arrays (ALMA)  become operational in  Cerro Chajnantor in  the Atacama
Desert plateau.

  We advocate more CI J=1--0 observations of similar objects at low as
well  as  high  redshifts  (where  CI lines  are  redshifted  to  more
transparent  atmospheric  windows)  in   order  to  further  test  the
potential of CI to trace bulk H$_2$ gas mass in galaxies.

\acknowledgments

We  thank the  referee for  comments on  the original  manuscript that
greatly improved the presentation of  the main thesis of this work. TRG
acknowledges  support from the  Danish Research  Council and  from the
European Union RTN network, POE

\clearpage

\begin{figure}
\includegraphics[angle=180,scale=1.00]{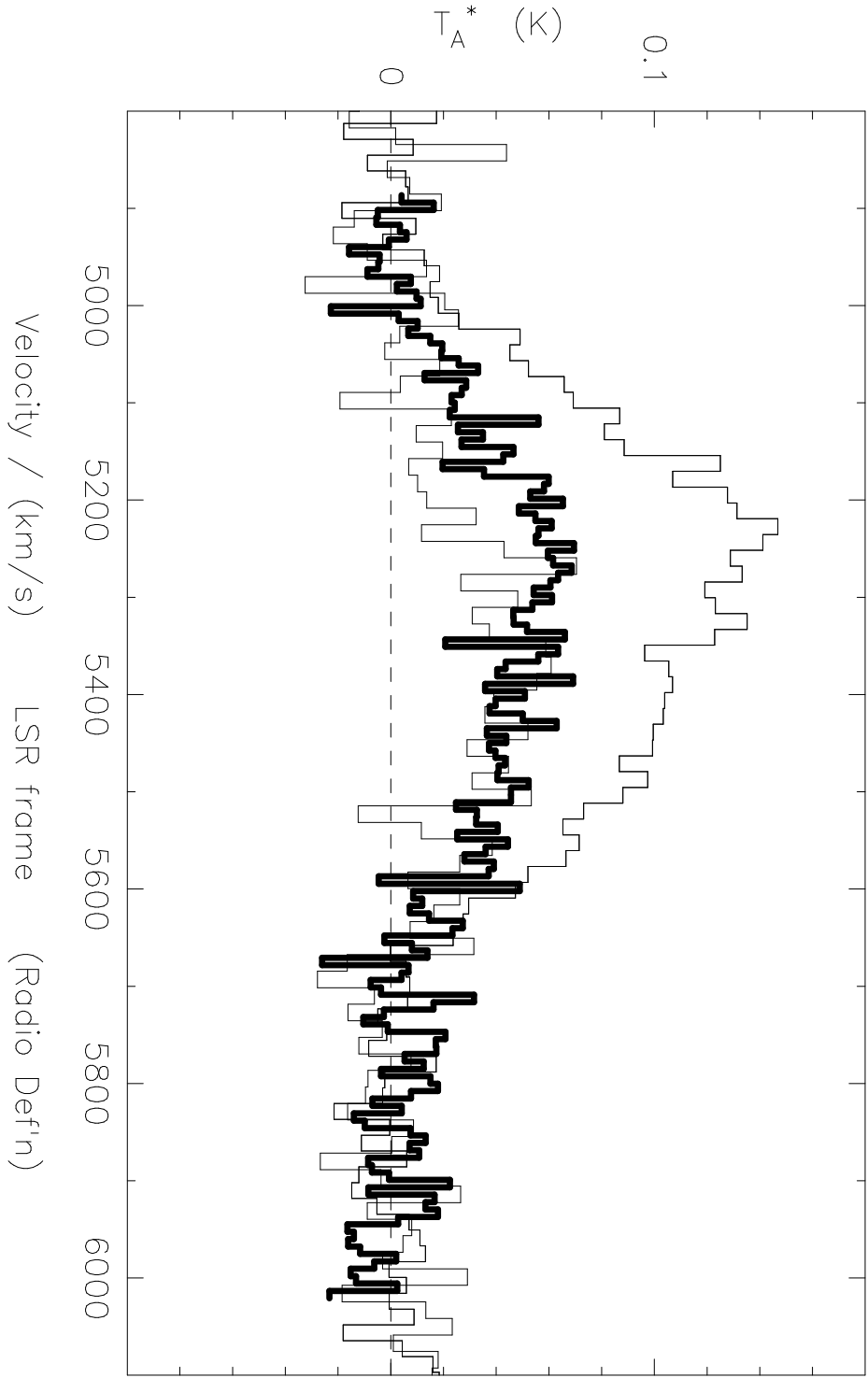}
\caption{Arp 220 ($\rm \alpha _{1950}: 15 ^h\ 32 ^{m}\ 46.73 ^s$, $\rm
\delta _{1950}: +23 ^{\circ}\ 40 ^{'}\ 07.36 ^{''})$. CI J=1--0 (thick
line), $  ^{12}$CO J=2--1 (strongest  spectral line), $\rm  10\times [
^{13}CO\  J=2-1]$  (thin  line).  The velocity-averaged  $  ^{12}$CO/$
^{13}$CO intensity  ratio is $\rm  R_{21}\sim 35$.  The  estimated rms
noise of the CI spectrum is $\rm \delta T^* _A = 10\ mK$ at a velocity
resolution of $\rm \Delta V_{chan}=7.6\ km\ s^{-1}$.}
\end{figure}

\begin{figure}
\includegraphics[angle=180,scale=1.00]{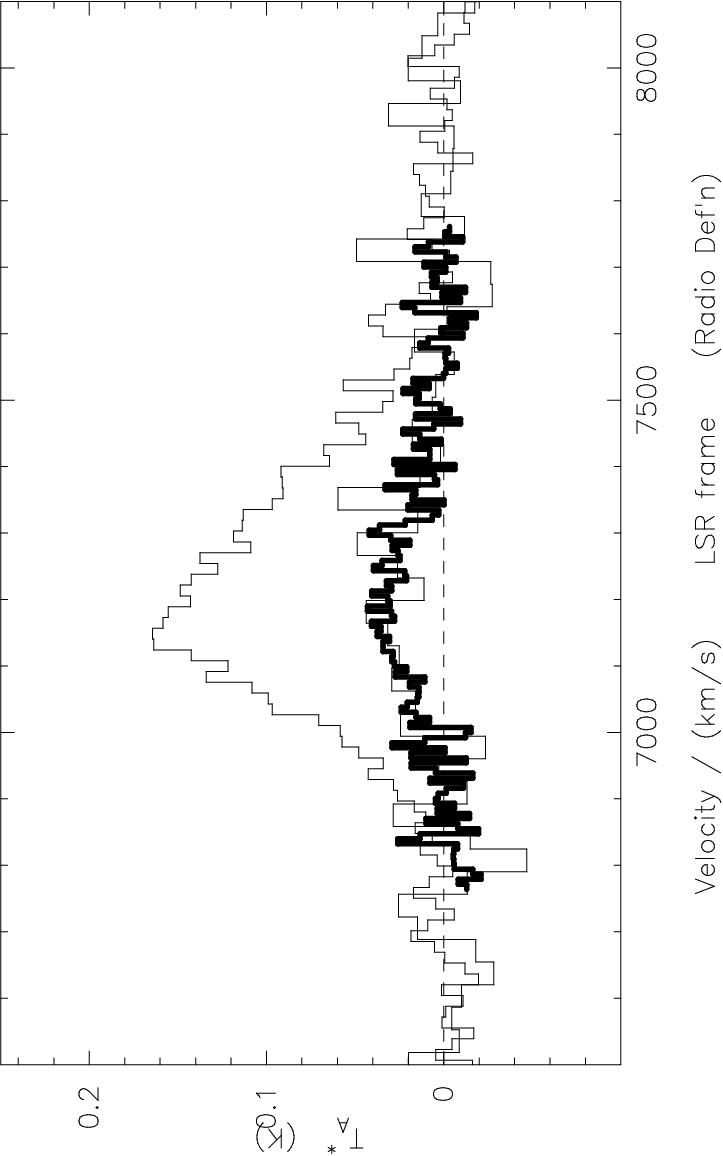}
\caption{NGC 6240 ($\rm \alpha _{1950}: 10 ^h\ 21 ^{m}\ 31.1 ^s$, $\rm
\delta _{1950}: +47 ^{\circ}\ 24  ^{'}\ 23.0 ^{''})$. CI J=1--0 (thick
line),  $ ^{12}$CO J=2--1  (strongest spectral  line) $\rm  10\times [
^{13}CO\  J=2-1]$  (thin line).   The  velocity-averaged $  ^{12}$CO/$
^{13}$CO intensity  ratio is $\rm  R_{21}\sim 50$.  The  estimated rms
noise of the CI spectrum is $\rm  \delta T^* _A = 8\ mK$ at a velocity
resolution of $\rm \Delta V_{chan}=7.6\ km\ s^{-1}$.}
\end{figure}

\newpage

\begin{deluxetable}{lcccccc}
\tablecolumns{8}
\tablewidth{0pc}
\tablecaption{Molecular gas mass estimates based on CI and CO}
\tablehead{
\colhead{Name} & \colhead{Redshift} &  \colhead{$\rm S_{CI}$} & \colhead{$(\rm T_k,  n)^b$} &
\colhead{$\rm Q_{10}$} & \colhead{$\rm M_{CI}(H_2)$} & \colhead{$\rm M_{CO}(H_2)^c$}\\
 &  &  (Jy km s$ ^{-1}$) & (K, cm$^{-3}$) &  & ($\times $10$^9$ M$_{\odot}$) & ($\times $10$^9$ M$_{\odot}$)}
\startdata
NGC 6240$ ^{\rm d}$ & 0.024 & 600$\pm $180 & (59, 10$^3$)   & 0.52 & $(6\pm 2)$ & 4 \\
Arp 220    & 0.018 & 1160$\pm $350 & (64, 10$^3$-10$^4$) & 0.47 & $(7\pm 2)$ & 5 \\
Mrk 231    & 0.042 & 200$\pm $55$ ^{\rm a}$  & (74, 3.5$\times $10$^3$) & 0.45 & $(7\pm 2) $ & 4 \\
NGC 6090$ ^{\rm e}$ & 0.029 & 270$\pm $55$ ^{\rm a}$  & (40, 10$^3$-10$^4$) & 0.48 & $(4\pm 1)$ & 6 \\
10565+2448 & 0.043 & 500$\pm $150$ ^{\rm a}$ & (60, 1.2$\times $10$^3$) & 0.51 & $(16\pm 5) $ & 4
\enddata
\\

\hspace*{-11.0cm} $ ^{\rm a}$ from Gerin \& Phillips 2000\\
\hspace*{-1.7cm }$ ^{\rm b}$  Solomon et al. 1997; Downes \& Solomon 1998
 (except NGC 6240, NGC 6090)\\
\hspace*{-5.5cm }$ ^{\rm c}$ Downes \& Solomon 1998 (except NGC 6240, NGC 6090)\\
\hspace*{0.1cm} $ ^{\rm d}$ $\rm M_{CO}(H_2)$ from Tacconi et al. 1999; $\rm (T_k, n)$
 from Large Velocity Gradient modeling of the \\
\hspace*{-4.6cm } line ratios reported by Tacconi et al.  and references therein.\\
\hspace*{-0.8cm} $ ^{\rm e}$ CO J=1--0 flux from Bryant \& Scoville 1999; Zhu et al 1999, 
 $\rm M_{CO}(H_2)$ deduced by \\ 
\hspace*{-4.8cm } using the X factor advocated by Downes \& Solomon 1998.
\end{deluxetable}

\end{document}